\documentclass[preprint, amsmath, amssymb, aps, prl]{revtex4-1}
\usepackage{graphicx}
\usepackage{bm}
\usepackage{amsmath}
\usepackage{epsfig}
\usepackage{braket}
\usepackage{tensor}
\usepackage{multirow}
\usepackage{mathrsfs}

\begin{document}
\title{Persistent directional current at equilibrium in non-reciprocal many-body near field electromagnetic heat transfer}
\author{Linxiao Zhu}
\email{lxzhu@stanford.edu}
\affiliation{Department of Applied Physics, Stanford University, Stanford, California 94305, USA}
\author{Shanhui Fan}
\email{shanhui@stanford.edu}
\affiliation{Department of Electrical Engineering, Ginzton Laboratory, Stanford University, Stanford, California 94305, USA}

\begin{abstract}
We consider the consequence of non-reciprocity in near-field heat transfer by studying systems consisting of magneto-optical nanoparticles.
We demonstrate that in thermal equilibrium, non-reciprocal many-body system in heat transfer can support a persistent directional heat current, without violating the second law of thermodynamics. 
Such a persistent directional heat current can not occur in reciprocal systems, and can only arise in many-body 
systems in heat transfer.
The use of non-reciprocity therefore points to a new regime of near-field heat transfer for the control of heat flow in the nanoscale.
\end{abstract} 

\maketitle
In most transport experiments, the presence of a current signifies that the system is away from equilibrium.
Electrical or mass current, for example, typically occurs when there is a voltage or density gradient.
On the other hand, the experimental discoveries of supercurrents in systems exhibiting superconductivity, superfluidity,
as well as the possibility for the existence of persistent directional current in ideal non-disordered systems exhibiting quantum Hall effects, represent some of the most intriguing effects in modern physics~\cite{kittel2004, Yoshioka2002}. 

\begin{figure}[htbp!]
\epsfig{file=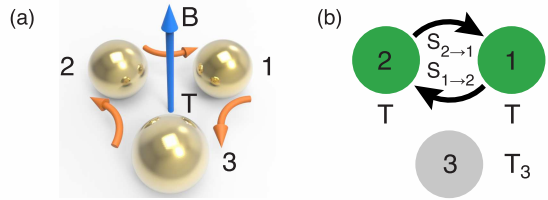}
\caption{\label{fig:persistent_schematic} (Color online).
(a) Persistent directional heat current in a many-body system, in thermal equilibrium of temperature $T$.
In our simulations, the three bodies consist of spheres made of magneto-optical materials forming an equilateral triangle, with a magnetic field applied perpendicular to the plane of the triangle.
(b) Heat transfer between bodies $1$ and $2$ at the same temperature $T$, while body $3$ is at temperature $T_3$.
}
\end{figure}

Heat currents are also commonly associated with non-equilibrium situations, such as in systems having a temperature gradient.
Nevertheless, inspired by the fundamental importance of supercurrents and persistent directional current in charge and mass transport,
in this Letter we consider persistent directional current at thermal equilibrium in heat transport.
The main results of our paper are schematically shown in Fig.~\ref{fig:persistent_schematic}, where we consider near-field electromagnetic heat exchange among three bodies having the same temperature.
We show that with broken reciprocity, the system in Fig.~\ref{fig:persistent_schematic} can exhibit such a persistent current, as represented by the significant difference in the directional heat flow between bodies $1$ and $2$,
even though all three bodies have the same temperature.
This is in strong contrast with reciprocal systems, where there cannot be any net heat flow between bodies $1$ and $2$ if they have the same temperature, 
independent of the temperature of the other bodies involved in the heat transfer. 
We also show that such persistent heat current is a genuine many-body effect in heat transfer - it can only occur when there are at least three bodies involved in the heat transfer. 

The study of near-field heat transfer is of substantial recent interests both from a fundamental perspective, since it highlights the importance of thermal electromagnetic fluctuations in nanoscale systems~\cite{Polder1971a, Loomis1994, Shchegrov2000, Joulain2005, Volokitin2007, Otey2010, Rodriguez2011a, Ben-Abdallah2014, Shen2009, Rousseau2009, Ottens2011, Kralik2012, St-gelais2014, Shi2015, Kim2015, Song2015, St-gelais2016, Song2016},
and also for practical applications such as energy conversion~\cite{Narayanaswamy2003, Basu2009b}, imaging~\cite{Kittel2005, DeWilde2006}, and radiative cooling~\cite{Guha2012, Chen2015}.
All previous theories on near-field heat transfer, however,
have only been applied to systems subject to the constraint of Lorentz reciprocity.
In particular, the developed theoretical formalism for near-field heat transfer makes extensive use of reciprocity~\cite{Kruger2012a, Messina2011a, Otey2014}.
Our works here highlight the important new opportunities for the control of near-field heat transfer by using non-reciprocal systems. 

We start by introducing the necessary conditions for achieving persistent directional heat current.
First of all, the system must break the Lorentz reciprocity. 
As an illustration, consider a system shown in 
Fig.~\ref{fig:persistent_schematic}b, consisting of bodies $1$, $2$ which have the same temperature $T$, and body $3$ which includes all other bodies involved in the heat exchange except bodies $1$ and $2$.
One can rigorously prove that, for a reciprocal system, $S_{1\to 2}=S_{2\to 1}$, 
regardless of the temperature of body $3$ (See a proof in Supplementary Material~\cite{SM2016}, which utilizes the formalism developed in Ref.~\cite{Kruger2012a}).
Here, $S_{i \to j}$ denotes the heat flow to body $j$ due to thermal excitation in body $i$.
Also, persistent heat current can only occur in a many-body system in heat transfer.
There needs to be at least three bodies involved in the heat transfer~\footnote{The minimal system can be three bodies involved in heat transfer, 
or two bodies with the environment.}.
For a system where heat transfer occurs entirely between only two bodies, 
in thermal equilibrium, the net heat flow between the two must be zero as dictated by the second law of thermodynamics, independent of whether the system is reciprocal or not.
On the other hand, the existence of a persistent heat current in the system shown in Fig.~\ref{fig:persistent_schematic}a does not violate the second law, since the total net heat flow into each body is zero when the three bodies have the same temperature, even in the presence of the persistent heat current.

\begin{figure}[h]
\epsfig{file=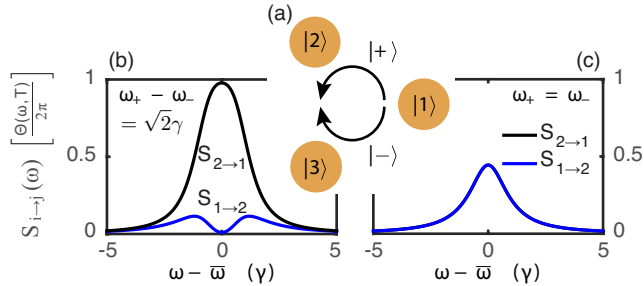}
\caption{\label{fig:GreenFunction} (Color online). 
(a) Schematic of a model for heat exchange among three bodies by way of states $\ket{+}$ and $\ket{-}$.   
Each body supports one mode.
(b) and (c) The heat transfer spectra for $S_{2\to 1}$ and $S_{1\to 2}$, from Eq.~\ref{eq:S_j_to_k}. $\overline{\omega} = (\omega_++\omega_-)/2$.
(b) Non-reciprocal case with $\omega_+-\omega_-=\sqrt{2}\gamma$.
(c) Reciprocal case with $\omega_+=\omega_-$.
}
\end{figure}

Motivated by the general theoretical consideration above, below we provide a simple analytic model that highlights the minimum requirement on the microscopic physics to support a persistent heat current. 
For this purpose we consider a three-body system with a $C_3$ symmetry in Fig.~\ref{fig:GreenFunction}a.
Each body supports a single mode $\ket{i}$, where $i=1$, $2$ and $3$.
Due to the rotational symmetry, the collective states of the system must have the forms:
$ \ket{+}=\frac{1}{\sqrt{3}} [1,e^{i\frac{2\pi}{3}}, e^{i\frac{4\pi}{3}}]^{T} $, 
$ \ket{-}=\frac{1}{\sqrt{3}} [1,e^{-i\frac{2\pi}{3}}, e^{-i\frac{4\pi}{3}}]^{T}$, 
$ \ket{\phi}=\frac{1}{\sqrt{3}} [1,1,1]^{T}$, 
as can be derived from considering the irreducible representations of the $C_3$ group. 
If the system is reciprocal, the frequencies of $\ket{+}$ and $\ket{-}$ states, denoted $\omega_+$ and $\omega_-$, respectively, must be equal. 
Breaking the reciprocity lifts the degeneracy between the two states.
But the form of the eigenstates $\ket{\pm}$ and $\ket{\phi}$ will not change as long as the $C_3$ symmetry is maintained. 
On the other hand, there is no symmetry constraint on the frequency $\omega_{\phi}$ of the state $\ket{\phi}$.
In what follows, we will consider the case where $\omega_{\phi}$ is significantly detuned from $\omega_{\pm}$, such that the heat flow is entirely carried by the states $\ket{+}$ and $\ket{-}$.
We assume the states $\ket{\pm}$ have the same internal decay rate $\gamma$.

The Green's function relating the field in body $k$ and the source in body $j$ is
\begin{eqnarray}
\mathscr{G}(k,j; \omega)&=& \sum_{P=\pm} \braket{k|P} \frac{1}{\omega- (\omega_P+i\gamma) } \braket{P|j}
= \frac{e^{i\frac{2\pi}{3}(k-j) }}{\omega-(\omega_++i\gamma) } 
+ \frac{e^{-i\frac{2\pi}{3}(k-j) }}{\omega-(\omega_-+i\gamma) } . \nonumber 
\end{eqnarray}
The spectral power density of heat transfer to body $k$ due to thermal noise source $n_j$ in body $j$ is 
\begin{equation}
\resizebox{0.94\hsize}{!}{$
 S_{j \to k} (\omega) =  \int_0 ^{\infty} d \omega^{\prime} 2 \gamma \left[ \mathscr{G}(k,j) \sqrt{2\gamma} n_j \right]^{\dagger}_{\omega}
 \left[ \mathscr{G}(k,j) \sqrt{2\gamma} n_j \right]_{\omega^{\prime}}
= \frac{\Theta(\omega,T_j)}{2 \pi}  \frac{4}{9}\left| \frac{e^{i\frac{2\pi}{3}(k-j) }}{1+i \frac{\omega-\omega_+}{\gamma} } 
+ \frac{e^{-i\frac{2\pi}{3}(k-j) }}{1+i \frac{\omega-\omega_-}{\gamma} } \right|^2,\label{eq:S_j_to_k} $}
\end{equation}
where $\langle n_j^{\ast}(\omega) n_{j^{'}}(\omega^{\prime}) \rangle = \frac{\Theta(\omega,T_j)}{2 \pi} \delta(\omega-\omega^{\prime}) \delta_{jj^{'}}$ from fluctuation dissipation theorem~\cite{Haus2000, Otey2010}.
Here, $\Theta(\omega,T) = \hbar \omega/ \left[  exp(\frac{\hbar \omega}{ k_B T})-1  \right]$.

We consider the thermal equilibrium case where all the three bodies have the same temperature $T$. 
When $\omega_+=\omega_-$, $S_{2\to 1}$ and $S_{1 \to 2}$ are identical (Fig.~\ref{fig:GreenFunction}c). 
In contrast, for the case with $\omega_+\neq\omega_-$, $S_{2\to 1}$ and $S_{1 \to 2}$ are no longer the same (Fig.~\ref{fig:GreenFunction}b). 
Accordingly, in thermal equilibrium, there is persistent heat current in this system. 
Here body $3$ has radiative heat exchange with the other two bodies, and is crucial for achieving persistent heat current.
For example, in the Supplementary Material~\cite{SM2016}, using a coupled mode theory model, we show that for a two-sphere system, even in the presence of the magnetic field, $S_{1 \to 2}(\omega)=S_{2 \to 1}(\omega)$.

In general, the strength of the persistent heat current between two bodies $i$ and $j$ can be characterized by the directionality of the heat flow between the two bodies, defined as 
\begin{equation*}
\eta_{ij}=|S_{i \to j}-S_{j \to i}|/min (S_{i \to j}, S_{j \to i} ).
\end{equation*}
In this system, for the spectrally integrated heat flux, the directionality is maximized to $\frac{2}{5}(2\sqrt{6}+3) \approx 316\%$ at $|\omega_+-\omega_-|=\sqrt{2} \gamma$.
We emphasize that we see such persistent directional heat current after spectral integration.
This is important as in most experiments of near field radiative heat transfer only spectrally-integrated heat flux can be measured. 
We also note that as the splitting between $\omega_+$ and $\omega_{-}$ increases beyond the optimal value, the directionality for spectrally integrated heat flux actually decreases. 
The directionality for the spectral density of heat current at a single frequency can reach infinity in this system, at $\omega=\frac{1}{2} (\omega_+ + \omega_-)$, when
$|\omega_+-\omega_-|=\frac{2}{\sqrt{3}} \gamma$.

The analytic model above relates the existence of persistent heat current, to the frequency-splitting of two collective counter-rotating states.
We now show that this concept can be implemented in a physical structure, as shown in Fig.~\ref{fig:persistent_schematic}a.
We consider three spheres with sub-micron diameters made of n-doped $InSb$, forming an equilateral triangle. 
An external $B$ field is applied in the vertical direction to break reciprocity. 
The relative permittivity tensor of $n$-$InSb$ in the presence of external $B$ field is:
$$
\Bar{\Bar{\epsilon}}=
\left( \epsilon_{\infty}+\epsilon_{inter}+\epsilon_{lattice}
\right)
\Bar{\Bar{I}}-
\frac{\omega_p^2}{ (\omega+i \Gamma)^2-\omega_c^2 }
\begin{bmatrix}
1 + i \frac{\Gamma}{\omega}  
& -i  \frac{\omega_c}{\omega}
& 0 
\\
 i \frac{\omega_c}{\omega}                          
 & 1 + i \frac{\Gamma}{\omega}  
 & 0 
 \\
 0
 & 0 
 &\frac{ (\omega+i \Gamma)^2-\omega_c^2 }{\omega (\omega+i \Gamma)} 
\end{bmatrix},
$$
where $\epsilon_{\infty}=15.68$ is the high-frequency permittivity, $\Gamma$ is the relaxation rate for free carrier, $\omega_c=eB/m^{\ast}$ is the cyclotron frequency, 
and $\omega_p=\sqrt{n_e e^2/(m^{\ast}\epsilon_0)}$ is the plasma frequency.
We use a doping concentration $n_e=1.36\times10^{19}~cm^{-3}$, 
for which $\Gamma=10^{12}~s^{-1}$ from experimental characterization~\cite{Law2014a}
and the effective electron mass is $m^{\ast}=0.08~m_e$ ($m_e$ is electron mass)~\cite{Byszewski1963, Law2014a}.
Here, in the permittivity tensor, we have included the contributions from free-carrier absorption, interband transition $\epsilon_{inter}$~\cite{Palik1985},
and lattice vibration $\epsilon_{lattice}$~\cite{Palik1985}.
\begin{figure}
\epsfig{file=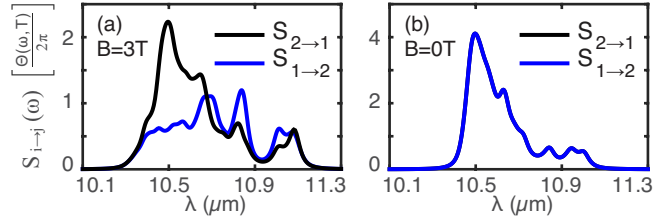}
\caption{\label{fig:3T} (Color online).
Heat transfer for the geometry shown in Fig.~\ref{fig:persistent_schematic}a. 
The spheres consist of $n$-$InSb$, with the same doping level.
Each sphere has a radius of $200~nm$, and the distance between the centers of two spheres is $500~nm$.
(a) and (b) The heat transfer spectra of $S_{1\to 2}$ and $S_{2 \to 1}$, from fluctuational electrodynamics. The system is in thermal equilibrium of $300~K$. 
(a) Non-reciprocal case with $B=3~T$. 
(b) Reciprocal case with $B=0~T$.
}
\end{figure}

In this system, the modes of individual spheres couple to form the kinds of pairs of circulating eigenmodes across three spheres as described in the analytic model discussed above.
With the application of the $B$ field, the frequencies of these pairs of circulating eigenmodes split, resulting in the persistent heat current.
For a detailed discussion of the modes involved in the persistent current, see Supplementary Material~\cite{SM2016}. 

To calculate the radiative heat transfer, we follow the scattering formalism based on fluctuational electrodynamics in Ref.~\onlinecite{Kruger2012a}, and extend the formalism for treating a general non-reciprocal many-body system.
In thermal equilibrium, based on fluctuation dissipation theorem (FDT)~\cite{Landau1980}, the correlator for electric fields for a general non-reciprocal system can be written as
$ \langle E(\mathbf{r}, \omega) E^{\dagger}(\mathbf{r}^{\prime}, \omega^{\prime}) \rangle
\equiv 2 \pi \delta(\omega-\omega^{\prime}) \langle E(\mathbf{r}) E^{\dagger}(\mathbf{r}^{\prime}) \rangle_{\omega}$, 
where
\begin{equation} \label{eq:EE_NR}
\langle E(\mathbf{r}) E^{\dagger}(\mathbf{r}^{\prime}) \rangle_{\omega}=
a(\omega,T) \frac{ G(\mathbf{r},\mathbf{r}^{\prime},\omega)- G^{\dagger}(\mathbf{r}^{\prime},\mathbf{r},\omega) }{2i}.
\end{equation}
Here, $G(\mathbf{r},\mathbf{r}^{\prime},\omega)$ is the dyadic Green's function relating a point dipole source at $\mathbf{r}^{\prime}$ to the electric field at $\mathbf{r}$,
and
$a(\omega,T)=\frac{4}{\pi}\omega \mu_0 \Theta(\omega,T)$. 
Using the field correlator resulting from thermal noise of body $1$, and a scattering formalism~\cite{Lippmann1950, Kruger2012a} for the Green's function, 
the spectral heat flux to body $2$ due to thermal noise in body $1$ is:
\begin{equation}
S_{1 \to 2}(\omega)=\frac{\Theta(\omega,T)}{2\pi} \cdot 4\cdot \mathbf{Tr} \left[ Q_2 W R_1W^{\dagger} \right], 
\end{equation} 
where $R_1=-\frac{1}{2}\left( T_1+T_1^{\dagger} \right) -T_1 T_1^{\dagger}$, 
$Q_2=-\frac{1}{2}\left( T_2+T_2^{\dagger} \right) -T_2^{\dagger} T_2$ and
\begin{equation*}
W= Z \left[ I-T_1 U_{13} T_3 U_{31} -T_1 \left( U_{12}+U_{13} T_{3} U_{32} \right) T_2 Z  \right]^{-1},
\end{equation*}
where $Z=\left(I-U_{23}T_3 U_{32} T_2 \right)^{-1} \left( U_{21}+U_{23}T_3 U_{31}\right)$.
Here, $T_i$ is the T matrix~\cite{Wittmann1988} for the isolated body $i$, 
and $U_{ji}$ is the conversion matrix connecting the wave modal basis for body $i$ and $j$.
For the spherical bodies considered, we use the vector spherical wave functions as the basis.
Unlike Refs.~\cite{Kruger2012a, Messina2011a, Otey2014}, for a non-reciprocal body $i$, $T_{i, [\{ l_1,m_1 \} , \{ l_2,m_2 \}]} \neq T_{i, [\{ l_2,-m_2 \} , \{ l_1,-m_1 \}]}$, where $l$ and $m$ are the quantum numbers for spherical waves.
We calculate the T matrix for the magneto-optical sphere, by constructing the eigenmodes inside the sphere using vector spherical wave functions, and then matching the boundary conditions~\cite{Li2011}.
We use a computationally-efficient recursive formalism to compute the conversion matrix using vector translation addition theorem~\cite{Chew1992, Chew1993}. 
We note it is well known that FDT applies only in the linear response regime.
Within the linear response regime, however, it does apply to systems with broken reciprocity and time-reversal symmetry, as pointed out for example in Refs.~\cite{Landau1980, Ben-Abdallah2016, Moncada-Villa2015}.
It is an interesting question whether near-field heat transfer in general is in the linear response regime~\cite{Perez-Madrid2013}.
Nevertheless, at present there is strong experimental support for the applicability of FDT in near field heat transfer.
A recent paper~\cite{Kim2015} shows excellent agreement between first-principles calculations from FDT and experiments. 

We show the calculated heat transfer spectra in Fig.~\ref{fig:3T}.
We first consider a system consisting of three deep-subwavelength spheres in near field. 
Each sphere has a radius of $200~nm$, and there is a $100~nm$ vacuum gap between each pair of two spheres. 
Figure~\ref{fig:3T}a shows the normalized heat flux spectra for $S_{2 \to 1}$ and $S_{1 \to 2}$.
We observe that the spectra $S_{1\to 2}$ and $S_{2 \to 1}$ are dramatically different at $B=3~T$, showing a strong directional heat flow.
At $\lambda=10.49~\mu m$, the directionality for the heat transfer between bodies $1$ and $2$ is as large as $276\%$.
After spectral integration over the whole thermal wavelength range from $1~\mu m$ to $100~\mu m$ at $T=300K$, the directionality for the heat transfer between bodies $1$ and $2$ is still as large as $30.5\%$, with $S_{2\to1}>S_{1\to 2}$. 
Moreover, due to the rotational symmetry of the system, $S_{2\to 1}=S_{3\to 2}=S_{1\to 3}$, and $S_{1\to 2}=S_{2\to 3}=S_{3\to 1}$.
Thus, in thermal equilibrium, there is strong persistent heat current in clockwise direction.
The direction for the persistent heat current can be flipped by reversing the external $B$ field.
The directionality for the spectrally integrated heat flux depends non-monotonously on the external $B$ field.
The directionality for the spectrally integrated heat flux will increase as $B$ increases, until peak at $58\%$ when $B=8.45T$.
As $B$ increases further, the directionality decreases~\cite{SM2016}.
This is consistent with our simple analytic model since the optimal directionality requires an optimal splitting of the modes involved and hence an optimal external magnetic field.
The scattering formalism agrees with a coupled mode theory model~\cite{SM2016}.
In contrast, for the scenario with $B=0~T$, $S_{2\to 1}$ and $S_{1 \to 2}$ are identical (Fig.~\ref{fig:3T}b).
We also note that the persistent heat current in Fig.~\ref{fig:3T}a is significant,
and its magnitude $|S_{2\to 1}-S_{1\to 2}|$ at equilibrium corresponds to the net heat flow between bodies $1$ and $2$ without external $B$ field (Fig.~\ref{fig:3T}b) at a temperature bias $|T_1-T_2|$ as large as $10.5~K$. 

\begin{figure}
\epsfig{file=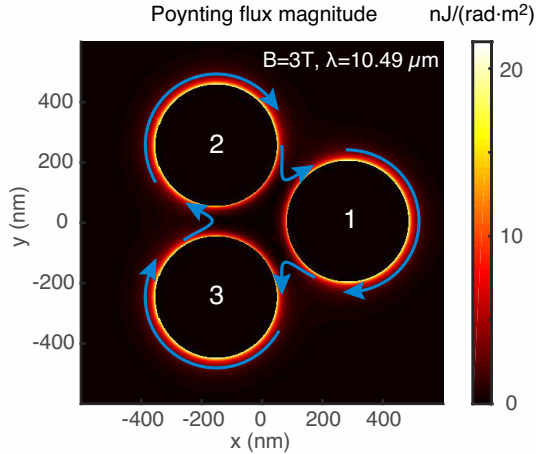}
\caption{\label{fig:small} (Color online).
Magnitude of Poynting flux, including all thermal noise sources in the spheres and environment, at $300~K$, $\lambda=10.49~\mu m$ and $B=3~T$. 
The geometry is the same as used in Fig.~\ref{fig:3T}.
The blue arrows denote the energy streamlines.
}
\end{figure}

To provide a visualization of  the persistent directional heat current, we show the profile of Poynting flux in the plane containing the sphere centers in Fig.~\ref{fig:small}. 
Here, the spheres and environment are at the same temperature $300~K$, and we plot the total Poynting flux resulting from thermal excitations in all the spheres and the environment. 
Figure~\ref{fig:small} shows the heat flux at $B=3~T$ and $\lambda=10.49~\mu m$,
and we observe intense persistent heat current in thermal equilibrium.
The magnitude of the Poynting flux is greatest at the surfaces of the spheres, 
and rapidly decays away from the spheres, 
indicating the near-field nature of the modes. 
In Fig.~\ref{fig:small}, we show as blue arrows the energy streamlines~\cite{Lee2007}, which trace the direction of the Poynting flux, at the gap regions and at the surfaces of the spheres.
The energy streamlines show that overall there is a persistent directional heat flux flowing between the spheres in clockwise direction, 
which agrees with the heat transfer spectra in Fig.~\ref{fig:3T}a.
Surrounding the spheres, the Poynting flux forms strong rotating pattern, implying a strong breaking of reciprocity. 
The demonstrated existence of persistent Poynting flux in thermal equilibrium does not violate the second law of thermodynamics. 
For each sphere, after integrating the energy flow over the whole surface area, the net power flow into each sphere is zero, as required by the second law.
While for ease of visualization we have plotted the Poynting flux at a single frequency, non-zero Poynting flux persists after spectral integration.
In contrast, for a reciprocal system ($B=0~T$) at equilibrium, we have observed that the heat flux of the whole system is strictly zero.

In the geometry used in Fig.~\ref{fig:3T}, both the dimensions of the spheres and gap sizes between the spheres are in the deep-subwavelength regime, and consequently the radiation to the far field is negligible.
That scenario corresponds to the simple analytic model, in which the particles are isolated from the surrounding environment. 
On the other hand, the persistent heat current can also exist in larger systems having significant radiative thermal exchanges with the environment. 
As an illustration, we consider a system where each sphere has a radius of $500~nm$ and there is a $1.2~\mu m$ vacuum gap between each pair of two spheres.
Remarkably, in this system, the directionality for spectrally integrated heat flux can reach as large as $61.2~\%$, at a relatively small magnetic field of $B=0.7~T$. 
In this system, the radiative heat transfer between each sphere and the environment is non-negligible, but reciprocal.

To further enhance the directionality of the persistent heat current, 
one may allow for different $B$ fields for each body. 
We note that capabilities for applying $B$ field with a strong spatial variation do exist, for example, in magnetic recording head~\cite{VanEk2000, Plumer2001}.
Our preliminary data also shows that with more bodies, the directionality may be further enhanced.
We note that while we have demonstrated the persistent heat current using $n$-$InSb$, 
the material choice is quite general, and 
many other semiconductors~\cite{Zawadzki1974, Raymond2001} such as $InAs$, $HgTe$, $Hg_{1-x}Cd_{x}Te$, $PbTe$, $PbSe$, $PbS$, or $GaAs$ which have low effective mass, and hence a large cyclotron frequency for a given external magnetic field, may be used for the purpose here.

We emphasize that the external magnetic field does not feed energy into the system.
The fluctuating current $j$ has an ensemble average of exactly zero in the presence of external $B$ field. Therefore, the ensemble averaged Lorentz force on such current, i.e. $\langle j \times B \rangle$, vanishes. 
Also, in our system, both the reciprocity and the time-reversal symmetry are broken.
The persistent current, however, requires non-reciprocity.
Time-reversal symmetry breaking alone is not sufficient for the persistent current.
Standard near-field heat transfer calculations consider structures described by a scalar dielectric function with an imaginary part.
These systems break time-reversal symmetry, but do not have persistent current since they are reciprocal.
We also note the effects shown here is fundamentally different from Ref.~\onlinecite{Ben-Abdallah2016}.
The persistent heat current shown here can exist even at thermal equilibrium. In contrast, the thermal Hall effect reported in Ref.~\onlinecite{Ben-Abdallah2016} is a non-equilibrium effect.
The thermal Hall effect, as measured in the temperature difference in the transverse direction, should vanish, in the absence of a longitudinal temperature gradient.

In summary, we have shown that there exists a persistent directional current in near-field heat transfer between dielectric objects described by non-reciprocal dielectric permittivity, even at equilibrium.
Such a persistent directional heat current represents a new regime of heat transfer - being in equilibrium such a current is not associated with entropy generation as far as thermal radiation is concerned.
As shown in Ref.~\cite{deGroot1984}, in heat flow in general the source for entropy generation vanishes when the temperature gradient vanishes, which is the case as we consider here. 
From a device point of view, the structures shown here can be an important component in the quest for controlling electromagnetic heat flow in nano-scale.
The circulating nature of the persistent directional heat current is reminiscent of thermal circulator~\cite{Li2015}.
It will also be important to explore this system away from thermal equilibrium. 
\begin{acknowledgments}
This work was supported by the DOE `Light-Material Interactions in Energy Conversion' Energy Frontier Research Center under Grant No. DE-SC0001293.
\end{acknowledgments}
\bibliographystyle{apsrev4-1}
%
\end{document}